\documentclass[pdflatex,sn-mathphys-ay,referee]{sn-jnl} 

\usepackage{graphicx}
\usepackage{amsmath,amssymb,amsfonts}%
\usepackage[title]{appendix}%

\usepackage[utf8]{inputenc} 
\usepackage[english]{babel} 

\usepackage{doi}
\usepackage{csquotes}

\hypersetup{
	pdftitle={Poincaré on Gibbs and on Probability in Statistical Mechanics}
	pdfsubject={physics.hist-ph, physics.class-ph},
	pdfauthor={Bruce D.~Popp},
	pdfkeywords={Poincaré, Gibbs,  many-body dynamical system, homoclinic points,  ontic probability, frequentist probability},
}

\begin{document}

\selectlanguage{english}

\title[Poincaré on Gibbs and Probability]{Poincaré on Gibbs and on Probability in Statistical Mechanics}

\author*[1]{\fnm{\href{https://orcid.org/0000-0001-9788-6466}{\includegraphics[scale=0.5]{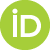}}\hspace{1mm}Bruce D.} \sur{Popp}}\email{BDPopp@Bien-Fait.com}
\affil*[1]{\orgname{Independent Scholar}, \orgaddress{\city{Laguna Niguel}, \state{CA}, \postcode{92677}, \country{USA}}}

\date{May 15, 2025}

\abstract{

This paper reviews a paper from 1906 by J. Henri Poincaré on statistical mechanics with a background in his earlier work and notable connections to J. Willard Gibbs. Poincaré's paper presents important ideas that are still relevant for understanding the need for probability in statistical mechanics. Poincaré understands the foundations of statistical mechanics as a many-body problem in analytical mechanics (reflecting his 1890 monograph on \emph{The Three-Body Problem and the Equations of Dynamics}) and possibly influenced by Gibbs independent development published in chapters in his 1902 book, \emph{Elementary Principles in Statistical Mechanics}. This dynamical systems approach of Poincaré and Gibbs provides great flexibility including applications to many systems besides gasses. This foundation benefits from close connections to Poincaré's earlier work. Notably, Poincaré had shown (e.g. in his study of non-linear oscillators) that Hamiltonian dynamical systems display sensitivity to initial conditions separating stable and unstable trajectories. In the first context it precludes proving the stability of orbits in the solar system, here it compels the use of ensembles of systems for which the probability is ontic and frequentist and does not have an \emph{a priori} value. Poincaré's key concepts relating to uncertain initial conditions, and fine- and coarse-grained entropy are presented for the readers' consideration. Poincaré and Gibbs clearly both wanted to say something about irreversibility, but came up short.
}


\maketitle

\section{ Introduction}

Reading a long-overlooked paper from 1906 by J. Henri Poincaré, \emph{Réflexions sur la théorie cinétique des gaz},\footnote{ There are similar titles and articles that need to be distinguished:   \citep{poincare_reflexions_1906} is the abstract appearing early in a volume of conference presentations and introduces his presentation appearing later in the same issue \citep{poincare_reflexions_1906-1}. That presentation is either nearly the same or exactly the same as the article \citep{poincare_reflexions_1906-2}. Since this later was published in a journal, it is the version considered here, exclusively. } with close connections to the book from 1902 by J. Willard Gibbs,\emph{ Elementary Principles in Statistical Mechanics},\footnote{ \citep{gibbs_elementary_1902} } reveals that Poincaré's 1906 paper suggests concepts with contemporary importance for the foundations of statistical mechanics.\footnote{ Readers looking for a review of the foundations of statistical mechanics would be well served by \citep{uffink_compendium_2007}. } It shows that Poincaré understood Gibbs's statistical mechanics as an application of many-body Hamiltonian dynamical systems; this is the perspective that Gibbs established in his first chapter. The essential insight is that conclusions from Poincaré's work over a decade earlier on the three-body problem in celestial mechanics, \emph{Sur le problème des trois corps et les équations de dynamique}\footnote{ Poincaré's work on dynamical systems is \emph{Sur le problème des trois corps et les équations de dynamique} \citep{poincare_sur_1890} and is available in translation as \emph{On the Three-Body Problem and the Equations of Dynamics} \citep{poincare_three-body_2017}. Also see \citep{poincare_methodes_1892} and two subsequent volumes; \citep{holmes_poincare_1990}. }, which introduced dynamical systems, have implications for the place of probability in statistical mechanics understood as a very-many-body system. 

The next section considers the immediate history of Poincaré's paper after its publication. Implicitly it suggests that the paper might have been overlooked because of negative responses in print by Jan Kroo and by Paul Ehrenfest and Tatiana Ehrenfest-Afanassjewa, and because of difficulty following Gibbs and Poincaré's exposition.

The following section provides a close summary of Poincaré's paper focused on those points and concepts most relevant to enduring questions about foundations of statistical mechanics in very-many-body systems, and omits his examples of non-equilibrium dynamical systems.

Last, Poincaré's suggestions from 1906 are discussed in relation to his recognition from over 15 years earlier of the importance to the three-body problem of acute sensitivity to initial conditions and consequent unstable trajectories. This is seen as justifying Gibbs's use of ensemble probabilities.

\section{ Background of Poincaré's Paper on Kinetic Theory}

In this, his final paper on kinetic theory of gases, Poincaré \citeyearpar{poincare_reflexions_1906-2} addressed three key points. In the abstract Poincaré critiqued embarrassing points with a lack of rigor leading to results having a paradoxical form and leading to contradictions, and also critiqued the concept of ``molekular geordnet,'' molecular order in German. In the body of the text, Poincaré additionally discussed changes in kinetic energy and entropy in an example of a non-equilibrium system. 

This section discusses some aspects of the context of his paper. In the following section important points from this article are summarized and the portion of his article, that will be considered further, is delimited.

\subsection{ Reception}

This 1906 paper seems to have gone mostly without comment at the time. I am only aware of two explicit contemporary citations: Jan Kroo \citeyearpar{kroo_uber_1911} in a footnote and Paul Ehrenfest and Tatiana Ehrenfest-Afanassjewa in their review article (\citep{ehrenfest_begriffliche_1911}; available in translation as \citep{ehrenfest_conceptual_1990}). 

In early 1911, Jan Kroo published an article arguing a point that he said Gibbs (in his \emph{Elementary Principles} \citep{gibbs_elementary_1902}) should have proven. At the time of publishing this article, Kroo was a graduate student in mathematics at Göttingen, Germany. With firm conviction, Kroo \citeyearpar{kroo_uber_1911} stated that a statistical ensemble of systems converges in general with time to a state of statistical equilibrium. He referred to this implausible assertion as the fundamental theorem of statistical mechanics and said that Gibbs relied on this theorem without explicitly stating or proving it. Kroo \citep[\S~2]{kroo_uber_1911} proposed to provide proofs of existence and stability of this state of statistical equilibrium. The attempted proof is seriously flawed by circular reasoning.

This 1911 article by Kroo is hardly worth mentioning except P.~Ehrenfest and T.~Ehrenfest-Afanassjewa \cite[pp. 71--2, \S~27]{ehrenfest_conceptual_1990} referred to it to critique both Gibbs and Poincaré. They---apparently without objecting to Kroo's proof---indicated that Kroo's correction was necessary because the ``treatment by Gibbs was incorrect.'' They supported this assertion with a cross-reference to their article \citep{ehrenfest-afanassjewa_bemerkung_1906} reviewing Gibbs's \emph{Elementary Principles}, where T.~Ehrenfest-Afanassjewa and P.~Ehrenfest referred to the subject of the first 10 chapters as, ``a peculiar mechanical analogy to the reversible processes of thermodynamics.'' In addition to mentioning Kroo's proof, they (still in section 27) also referred to a footnote \citep[n.~1, p.~919]{kroo_uber_1911}, where Kroo said that Poincaré's example with asteroids was examined in an imperfect way. Kroo offers no amplification and the Ehrenfests use the footnote to dismiss Poincaré's 1906 paper without further discussion.

\subsection{ Irreversibility and Entropy}

From Poincaré's other work \cite[e.g.][]{poincare_entropy_1903}, we know he understood the second law of thermodynamics as a consequence of induction from experience. He stated that transitions between two accessible states with equal energy only occur in one direction. For Poincaré, entropy is then some quantity that increases during this transition and he understands this as requiring a state-variable that monotonically increases as the system passes between two states in the allowed direction; this definition is not unique, since it leaves the initial value and a multiplicative constant undefined.\footnote{ In contrast Boltzmann's H theorem allows entropy to increase or decrease such that the total change is a net zero but assumes that it is extremely improbable that we will see a part of the H curve where entropy decreases. Why should we accept that improbability instead of the alternative, insisting that entropy is monotonic? } Increasing entropy is then a consequence of the observational fact of irreversibility.

Concern about explaining irreversibility in mechanical analogs of thermodynamics is commonly traced to Josef Loschmidt in 1876. Ludwig Boltzmann made a concerted effort to overcome his objection. Twenty years later Ernst Zermelo and Boltzmann engaged in a famous dispute about irreversibility. Zermelo based his arguments in Poincaré's work on dynamical systems, specifically his recurrence theorem. He saw the determinism inherent in the laws of analytical mechanics as an insurmountable barrier to using them to explain irreversibility in thermodynamics and did not see probability theory as a path around this barrier. A further ten years after the polemic engulfing Zermelo and Boltzmann, indeed after the death of Boltzmann, Zermelo, in his review of Gibbs's \emph{Statistical Mechanics} (\citep{zermelo_notizen_1906}; reprinted and translated in \citep[II:571--592]{zermelo_ernst_2013}), insisted (see page 589) that to explain thermodynamic irreversibility Hamilton's equations needed to be replaced with ``differential equations already containing the principle of irreversibility in themselves.'' Zermelo did not address the subject again. 

Zermelo's call for replacement differential equations seems to have gone unanswered. Therefore, there appears to be a continuing need for a better, or at least more widely accepted, explanation of macroscopic irreversibility based on reversible, analytic laws of motion of the constituent bodies.

\subsection{ Analytical Mechanics}

The origin of Poincaré's 1906 article is perhaps best understood as his effort to expand on and clarify points from J. Willard Gibbs's book,\emph{ Elementary Principles in Statistical Mechanics}, and Ludwig Boltzmann's concept of molecular disorder. Poincaré had a very positive reaction to the work of Gibbs; previously, Poincaré---and also Pierre Duhem, see, for example \citep{duhem_potentiel_1886}---had been an early reader of Gibbs's work on thermodynamics. Gibbs's book, \emph{Statistical Mechanics,} and Poincaré's book,\emph{ On the Three Body Problem and the Equations of Dynamics}, from 12 years earlier can profitably be read and understood as companion volumes. They both built on a solid foundation in Newtonian or analytical mechanics. They required that the forces experienced by the bodies be expressible by a potential that is a differentiable function of the position and momentum of the bodies. If the force is gravitation, the interactions occur continuously, at great distances, and without collisions. In the case of atoms and molecules, the forces are electromagnetic and arise from short-range interactions between the higher order multipole moments of net-neutral bodies. 

Following variously \citep[p.~3--5]{gibbs_elementary_1902}; \citep[\S~10]{tolman_principles_1979}; or, with slightly different notation, \citep[p.~7--43]{poincare_three-body_2017}, the Hamiltonian or canonical equations of motion for $f$ degrees of freedom are the $2f$ linear differential equations:
\[ \begin{array}{ccc}
\frac{\mathrm{d}q_i}{\mathrm{d}t}=\frac{\partial H}{\partial p_i} & \mathrm{and} & \frac{\mathrm{d}p_i}{\mathrm{d}t}=-\frac{\partial H}{\partial q_i} \end{array}
.\] 

The $q_i$ and $p_i$ are referred to as conjugate variables, with $q_i$ being the generalized coordinate and $p_i$ the conjugate momentum for the $i$${}^{th}$ degree of freedom. Their time derivatives are commonly written as ${\dot{q}}_i$ and ${\dot{p}}_i$.

The Hamiltonian $H$, which in the scenario considered by Poincaré and Gibbs does not explicitly depend on time, $t$, is
\[H\mathrm{=}H\left(q_1,\cdots q_f,p_1,\cdots p_f\right)\mathrm{=}T\mathrm{+}V,\] 
where $T$ is the kinetic energy and is generally a function of the ${p_i}^2$, and where $V$ is the potential energy and is a function of the coordinates $q_i$. The potential energy of a system where each body has one degree of freedom may depend on some or even all the other coordinates. The Hamiltonian is thus the total energy.

There is a distinction to be made here between the foundation that Gibbs and Poincaré developed and built on, and the earlier foundation used by James Clerk Maxwell and Ludwig Boltzmann. This distinction is relevant and even important in understanding and comparing Gibbs and Boltzmann entropies. This distinction is known \citep[see for example: ][p.~xxvii]{pathria_statistical_2022}, but a reminder is often necessary.\footnote{ E.~T. Jaynes gave the need for this reminder as justification for publishing his note \citeyearpar{jaynes_gibbs_1965}. }

The work of Maxwell and Boltzmann depended on kinematic assumptions about the collisions between gas atoms or molecules. These assumptions started with instantaneous, elastic collisions between pairs of hard spheres. Much has been written about Maxwell and Boltzmann's consideration of collision numbers, velocity fields and the like; collectively they are sometimes, perhaps incorrectly, referred to as the \emph{Stoßzahlansatz} following Paul Ehrenfest and Tatiana Ehrenfest-Afanassjewa. Their use of kinematic assumptions does not justify the description ``mechanics'' first used by Gibbs in 1884.\footnote{ The abstract of a presentation, \emph{On the Fundamental Formula of Statistical Mechanics, with Applications to Astronomy and Thermodynamics}, to a meeting in 1884 of the American Association for the Advancement of Science are included in Gibbs's collected works \cite[2:16]{gibbs_scientific_1906} but there is apparently no record of the presentation. } 

With their shared foundation in analytical mechanics, Poincaré and Gibbs could both prove theorems about invariance of phase space volume (extension in phase) and about recurrence, and could separately develop subject-specific applications. For Poincaré this included the three-body problem. For Gibbs this included statistics of ensembles. The generality of the foundation in analytical mechanics that they built on allowed Gibbs to develop a theory without concern for difficulties like the number of degrees of freedom in a diatomic gas molecule (a widely quoted concern from his preface) and then to use the theory to treat gas molecules (his final chapter). It also allowed Poincaré to develop applications to undamped non-linear oscillators (\citep[p.~155--58]{holmes_poincare_1990}; e.g. undamped Duffing's equation, \citep[p.~103, 157]{poincare_three-body_2017}),celestial mechanics (\citep[and two subsequent volumes]{poincare_methodes_1892}), globular clusters (\citep{poincare_voie_1906}, available in translation as \citep{poincare_milky_1906}; \citep{popp_early_2022}) and quantized, radiating harmonic oscillators \citep{poincare_sur_1912} as specific cases. These examples involve bodies with measurable dimensions and with interactions occurring over a few atomic radii, many astronomical units or even many parsecs.

This distinction between building on analytical mechanics and building on assumptions about collisions is meaningful beyond the range of specific subjects that can be accessed. The approach used by Gibbs and Poincaré can address molecules that are not monoatomic (i.e. have only translational degrees of freedom); for example, it can address polyatomic molecules with rotational or vibrational degrees of freedom, since each different degree of freedom, $i$, has its own generalized coordinate, $q_i$, and conjugate momentum, $p_i$. Therefore indexes $i$ through $i+5$, correspond, for example, to a single diatomic molecule with 6 degrees of freedom (i.e. three translational, two rotational and one vibrational). This is the physical reason, as Jaynes \cite[p.~391]{jaynes_gibbs_1965} noted, ``The Gibbs formula gives the correct entropy, as defined in phenomenological thermodynamics, while the Boltzmann $H$ expression is correct only in the case of an ideal gas.''

\section{ Poincaré's 1906 Paper}

This section deals with the content presented in Poincaré's paper. The first subsection ({\S}3.1) provides a high-level presentation of the issues and concepts covered by Poincaré. After a transition relating to Gibbs and Poincaré connecting over mixing, the final two subsections ({\S}3.3 and {\S}3.4) enter into details from Poincaré's article on initial conditions, and fine and coarse graining. The remaining sections of Poincaré's paper concerning non-equilibrium configurations is not considered here.

\subsection{ Key Points}

In his final paper on kinetic theory of gases, Poincaré \citeyearpar{poincare_reflexions_1906-2} addressed three subjects. In the abstract Poincaré \citeyearpar{poincare_reflexions_1906-2} critiqued embarrassing points with a lack of rigor leading to results having a paradoxical form and leading to contradictions, and also critiqued the concept of ``molekular geordnet,'' molecular order in German. In the body of the text but not mentioned in the abstract, Poincaré additionally discussed changes in kinetic energy and entropy in a model non-equilibrium system.

\textbf{The first point} of these embarrassing points, the one that Poincaré says embarrassed him the most, was the continuing failure to prove that entropy increases.\footnote{ Note that Poincaré in this paragraph does write, ``l'entropie va en diminuant.'' This may immediately confuse the reader expecting to read, ``the entropy increases.'' The definition of entropy in equation 3 on the second page, consistent with the definition in Gibbs, does not have the factor $-k$ conventionally included in the definition: $\int{P{\mathrm{ln} P }{\mathrm{d}}^nq{\mathrm{d}}^np}$. Poincaré has thus locally redefined the term entropy to be its negative (sometimes called negentropy by other authors). Bumstead \cite[p.~13]{bumstead_variation_1904} similarly included an entirely analogous footnote in his discussion of Gibbs's \emph{Statistical Mechanics}. } Indeed, careful consideration of Gibbs's definition of entropy has a prominent and essential position in \citeyearpar{poincare_reflexions_1906-2}. Poincaré asked whether, with the definition and discussion from \cite[chap. XII]{gibbs_elementary_1902}), entropy does increase. This is the point relevant to the present paper.

To investigate this point, Poincaré introduced \emph{two key concepts}: the discontinuous and continuous hypotheses (uncertainty of initial conditions), and fine-grained and coarse-grained entropy. These concepts are each discussed in their own subsections below. As will be seen they are respectively related to the origin of the need for probability and statistics, and evaluation of limits as volumes in real or phase space approach zero. Poincaré understood the need for statistics as due to uncertainty in the initial configuration in phase space of a system of molecules. The evaluation of limits expands on and provides examples related to subjects and concepts connected with mixing coloring in water. 

\textbf{The second point} is related to order and disorder in the distribution of molecules in phase space. This German term, \emph{molekular} \emph{geordnet}, appears in \citep[p.~21]{boltzmann_vorlesungen_1896} and in its translation into English appears in \citep[p.~17--18]{boltzmann_lectures_1995}. Poincaré did not indicate how he understood molecular disorder and he does not mention Boltzmann, so it is not clear whether he was trying to make a specific response to Boltzmann. Molecular disorder is not considered further in Poincaré's paper.

Here Poincaré introduced a \emph{third key concept}: apparent and latent organization. Poincaré relates organization to a distinctive spatial arrangement of asteroids, and the recurrence of that arrangement. The concept of organization is not well-defined and since Poincaré makes no further mention of \emph{geordnet}, there is nothing to suggest how he understands the relation between organization and disorder. 

Poincaré's example of apparent organization involves asteroids in Keplerian orbits around a central body. He indicates that an arrangement of these asteroids in conjunction (meaning at the same longitude) shows apparent organization. In the absence of organization, as Poincaré (\citeyear{poincare_three-body_2017}, p.~378) writes, ``the distribution will seem uniform and without systematic trend to a coarse view.'' This example makes clear the importance of the observer's assessment of the configuration.

For comparison, there are better treatments of organization and disorder from the 1900s in works on probability by Poincaré, Émile Borel and Jacques Hadamard. They discuss playing cards and ask how to recognize a particular configuration of a shuffled deck of cards as either disordered or organized. To answer whether a deck of cards is properly shuffled, they ask about the probability of certain sequences of cards turning up as individual cards are dealt sequentially. With hindsight, a parallel can be recognized between the cards being dealt from a deck and the signals arriving on a communication channel. 

\textbf{The third point} is the condition from Gibbs that changes in kinetic energy and entropy need to be evaluated near equilibrium. Poincaré asks if this condition is indeed necessary and emphasizes the importance to him of relaxing the condition. In an extended example, he considers a model---he calls it a one-dimensional gas---that he states is subject to the gravitational potential of a disrupting body that is brought in and removed at different times. He considers the kinetic energy and entropy of the one-dimensional gas first sufficiently long after the arrival of the disrupting body that the gas has reached equilibrium, and then a short time (before reaching equilibrium) after the arrival. The model is limited and has properties that Poincaré calls ``paradoxical;'' Poincaré uses the model to show that in non-equilibrium configurations (a short time after removing or introducing the disrupting body) the model undergoes changes in kinetic energy and entropy that are consistent with the values reached in equilibrium.

\textbf{The scope of the remaining discussion }in the present paper is limited to the concepts introduced in connection with the first point, above: uncertainty of initial conditions, and fine-grained and coarse-grained entropy.

\subsection{ Mixing, an Interlude}

One of those clear connections---a point in Gibbs's \emph{Principles of Statistical Mechanics} to which Poincaré responded with interest---was an analogy of mixing color into water. \citep[p.~144]{gibbs_elementary_1902} wrote, ``Let us suppose the liquid to contain a certain amount of coloring matter.'' Even before his 1906 paper, Poincaré had responded to Gibbs's example of mixing in the following paragraph from \citep[p.~308]{poincare_letat_1904}\footnote{ This paragraph is in a speech delivered to the International Congress or Arts and Science, St. Louis, September~1904. The speech is also available in English in both \citep{poincare_principles_1905} and \citep{poincare_present_1906}, but note that the translation there swaps reversible for irreversible and has lesser inaccuracies too. The translation here is my own.}:

Should a drop of wine fall in a glass of water: whatever the liquid's internal law of motion, we will soon see it colored a pinkish hue and, from that moment, you can shake the container however you want, the wine and water no longer seem able to separate. Thus, here is what would be the model of the irreversible physical phenomenon; hiding a grain of barley in a cup of wheat is easy to do. Finding it again and taking it out of there is nearly impossible. Maxwell and Boltzmann explained all that, but the one who saw most clearly, in a book not read enough because it is a bit difficult to read, is Gibbs, in his \emph{Elementary Principles in Statistical Mechanics}.

Poincaré, being thoroughly French, personalized this example by changing the coloring material to wine. The underlying issue is understanding irreversibility. A decade earlier Poincaré \citeyearpar{poincare_mecanisme_1893} had suggested an illustration of irreversibility and mixing in which a grain of barley is placed in a sack of wheat. The close connection between the two independent statements about mixing grains and liquids is striking. Poincaré's choice of mixing elements (grains) of clearly finite size is interesting in light of his subsequent discussion of coarse graining.

And there are connections where the nature is less clear. \citeyearpar{gibbs_elementary_1902} started Chapter XII with a statement of Poincaré's recurrence theorem from his work on dynamical systems and provided his own proof. Had Gibbs read earlier work by Poincaré in dynamical systems? Is Gibbs's proof an independent rederivation? We don't know. Hadamard \cite[p.~196]{hadamard_book_1906} saw a broader influence of Poincaré in the work of Gibbs. Hadamard stated that Gibbs ``owes Mr.~Poincaré the development of the principles of the general theory of differential equations and analytic mechanics.'' As an example, Hadamard noted the concept of \emph{integral invariant} and its applications. Hence, he noted that phase-space volume, or extension-in-phase, is an integral invariant and continues with how Gibbs extended this by showing that extension-in-phase is invariant under change of the coordinate system used.

However, the \emph{style} of Gibbs's approach to entropy in Chapter XII and of Poincaré's approach in \citeyearpar{poincare_reflexions_1906-2} are notably different.

In commentary on Gibbs's \emph{Statistical Mechanics} several reviewers, and notably Samuel Burbury \citeyearpar{burbury_variation_1903}, commented on the reserve, hesitancy and insufficient rigor in Chapter XII (as compared, for example, to Chapter I). Martin Klein \cite[p.~15]{klein_physics_1990} suggested that this attitude was related to Gibbs's dissatisfaction with the analysis of irreversibility.

His own discussion of the irreversible approach to equilibrium is set forth in Chapter XII entitled ``On the Motion of Systems and Ensembles of Systems Through Long Periods of Time.'' Gibbs treated it as a mixing process in phase space, but his guarded language and the absence of equations in this chapter suggest that he may not have been completely satisfied with this analysis of irreversibility.

Poincaré, however, proceeded boldly and concluded his paper with a note of confidence that he had found a resolution, or a path to a resolution. This confidence rested on a set of concepts and a detailed example. The first two concepts are durable. The third concept, organization, related to probability and information, is explained but not defined and needed to be sent back for further work.

\subsection{ Uncertainty of Initial Conditions}

Poincaré started his presentation of the concepts using the probability function (that Gibbs also refers to as $\eta $ and Poincaré refers to as $P$) and an infinitesimal extension-in-phase $\mathrm{d}\tau $ (where $\mathrm{d}\tau =\mathrm{d}x_1\mathrm{d}x_2\cdots \mathrm{d}x_n$, in Poincaré's notation, or $\mathrm{d}\tau =\mathrm{d}q_1\mathrm{d}q_2\cdots \mathrm{d}q_m\mathrm{d}p_1\mathrm{d}p_2\cdots \mathrm{d}p_m$, in Gibbs's notation, where $n=2m$). The phase space volume or extension-in-phase of the system is then the n-fold integral
\[\int{\mathrm{d}x_1\mathrm{d}x_2\cdots \mathrm{d}x_n}=\int{\mathrm{d}\tau }\] 
extended over the domain.

After these definitions, Poincaré \cite[p.~371]{poincare_reflexions_1906-2}) called for a clear understanding of the probability function and the following integrals:
\[\int{P\mathrm{d}\tau },\int{P{\mathrm{log} P }\mathrm{d}\tau },\int{\varphi P\mathrm{d}\tau }.\] 

The first issue to be understood here was the origin of the probability in the probability function and therefore in statistical mechanics. Poincaré identified two possibilities that he refers to as the \emph{discontinuous} and \emph{continuous} hypotheses, corresponding to certainty and uncertainty of the initial conditions of a system.

He started by assuming that there are many similar systems (Poincaré wrote \emph{semblable}). Each system could for example be a molecule. Or the system could be a gas with many molecules, where the many systems are only different because the gas molecules are arbitrarily permuted. 

The \emph{discontinuous hypothesis}, Poincaré explains, further assumes that the \emph{initial conditions of motion are fully known} for each system. The immediate consequence of this hypothesis, since Poincaré is working with Hamiltonian equations of dynamics based on deterministic analytical mechanics (and not working with assumptions about collision frequencies), is that the conditions of motion are also fully known at any subsequent instant. The probability that a system is in a given domain at a given instant is then the ratio of the number of systems in this domain at that moment to the total number of systems. Under the \emph{discontinuous hypothesis}, there is no uncertainty and no need for probability since the laws of mechanics are deterministic, and the initial conditions and therefore all subsequent conditions are all fully known. 

The \emph{continuous hypothesis}, as Poincaré explains, takes the contrasting assumption that the \emph{initial conditions of motion of the system are }not\emph{ fully known}. The conditions are likewise uncertain at any subsequent instant. The probability that the system will be in a very small volume in phase space then reflects the ongoing uncertainty of the condition of the system. 

It was then a matter for Poincaré to choose one of these exclusive hypotheses that he presented. The two hypotheses lead to divergent conclusions for the calculation of the entropy when dividing phase space into smaller volumes for the calculation of the entropy. As the volumes become smaller, but not yet infinitesimal, under the discontinuous hypothesis there will be volumes where it is fully known whether there is a particle, and those volumes will make a zero or infinite contribution to the entropy. The sum over all volumes will result in an entropy that is infinite (in magnitude). In contrast, under the continuous hypothesis, the calculated entropy remains finite. Poincaré therefore adopted the continuous hypothesis. 

Poincaré, by choosing the \emph{continuous hypothesis}, set aside the assumption underlying the discontinuous hypothesis, perfect knowledge of the initial conditions. He \cite[p.~371]{poincare_reflexions_1906-2} stated that, ``For each system, the initial conditions of motion are not fully known and {\dots} we can only evaluate the probability that at the origin of time the system considered is located in a certain domain; all that we can do will be to evaluate the probability that for that instant the system will be located in a given very small domain.'' Probability, statistics, is required because the initial conditions are not completely known. Poincaré then generalized from a single system with uncertain initial conditions and a probability $p\mathrm{d}\tau $ of being in an extension in phase $\mathrm{d}\tau $ to ``many [N] similar systems, where the probability that a system is in the extension of phase $\mathrm{d}\tau $ will be $\frac{\sum{p}}{N}\mathrm{d}\tau $.''

By making this connection to uncertainty of initial conditions, Poincaré has connected probability in Gibbs's \emph{Statistical Mechanics} with an important property of Hamiltonian dynamical systems that he discovered in his study of the three-body problem and the equations of dynamics \citep{poincare_sur_1890}. Small differences in the initial conditions near homoclinic points can separate stable and unstable trajectories and result in large differences in subsequent behavior over sufficient time.${}^{ }$\footnote{ For an example of the consequences of even minor uncertainty, consider the Proposition and discussion by Philip Holmes on page 154 in \citeyearpar{holmes_poincare_1990}.} This uncertainty, and its consequences, are a feature of dynamical systems now commonly referred to as chaos. Poincaré had shown that the stability of the solar system cannot be proven and here shown that the use of probability in statistical mechanics cannot be avoided because even minor changes of the initial conditions can make the difference between stable and unstable trajectories.

\subsection{ Fine-Grained and Coarse-Grained Entropy}

There is however another difficulty in the calculation of the entropy in the approach to infinitesimal volumes in phase space. The product of the probability function and its log, appearing in the calculation of Gibbs entropy, even though finite and continuous, could vary widely over a small volume, with the consequence that the value of the product at the center of the small volume times the value of the volume could be different than the integral of the product over that volume. This is a mathematical statement closely related to Gibbs's analogy of coloring matter in a liquid. A small volume of space in the liquid, after mixing for some time, may contain smaller volumes with coloring matter and other volumes without any. The calculation of the density of the coloring matter then depends on: 1) where the quantity of coloring matter is measured in the volume, and 2) whether the volume is divided into smaller volumes for the calculation.

Take a domain of phase space with small volume $\delta $ and let $\mathrm{\Pi }$ be the probability that the system is in that domain. Consider the following sums:
\[\sum{\mathrm{\Pi }\delta }\sum{\mathrm{\Pi }{\mathrm{log} \mathrm{\Pi } }\delta }\sum{\mathrm{\varphi }\mathrm{\Pi }\delta }.\] 
Subject to the continuous hypothesis, in the limit of $\delta $ approaching $\mathrm{d}\tau $, then $\mathrm{\Pi }$ will approach $P$. These sums approach the integrals discussed above:
\[\int{P\mathrm{d}\tau },\int{P{\mathrm{log} P }\mathrm{d}\tau },\int{\varphi P\mathrm{d}\tau }.\] 

It is in this context of taking the limit, with reference to his example of mixing coloring, that \cite[p.~146]{gibbs_elementary_1902}) refers to, ``the fineness of our method of evaluating the density.'' Gibbs is thus advocating stopping the approach of $\delta $ at a finite value before reaching the mathematical, infinitesimal limit. This is a point with which Zermelo disagreed in his review of Gibbs's \emph{Statistical Mechanics}. Zermelo argued that evaluating these sums in the limit of $\delta $ approaching zero is mathematically valid and stopping at a finite value is unjustified (at least without building a theory on that assumption).

Poincaré \citep[p.~373]{poincare_reflexions_1906-2} distinguishes two cases: the result of taking the limit of $\sum{\mathrm{\Pi }{\mathrm{log} \mathrm{\Pi } }\delta }$ for vanishing $\delta $ is \emph{fine-grained entropy}; the result of stopping at a finite $\delta $ is \emph{coarse-grained entropy}. 

Poincaré's 1906 appears to be the first use of the terms fine- and coarse-grained. The review article by Paul Ehrenfest and Tatiana Ehrenfest-Afanassjewa \cite[p.~52, \S~23]{ehrenfest_conceptual_1990}, used the same terms and appeared five years later. The concept (without the terms) has its origin with Gibbs.

Zermelo's argument, restated, is that only the calculation of fine-grained entropy is valid.

Coarse-grained entropy is what Gibbs proposed and supported with his discussion of mixing coloring and water. Poincaré \cite[p.~373]{poincare_reflexions_1906-2} set the finite value of $\delta $ for coarse-grained entropy as a size that was ``small enough such that our usual means of investigation no longer allow us to distinguish two phases inside a single domain.''\footnote{ However contrast this statement from 1906 to his statement 13 years earlier in Poincaré \citeyearpar{poincare_mecanisme_1893}, ``The apparent irreversibility of natural phenomena could in the same way be because molecules are too small and too many for the coarseness of our senses.'' I have wondered whether there is an explanation for irreversibility in well known, non-deterministic behavior of dynamical systems; I have not been able to defend an opinion to my satisfaction.} Henry A. Bumstead \cite[p.~9]{bumstead_variation_1904} restated Gibbs posthumously, ``It is because we are unable, owing to the finiteness of our perceptions, to recognize very small differences of phase, just as we are unable to recognize very small differences of position in the analogous case of the liquid.'' At this point the obvious question for Gibbs and Poincaré is, why are subjective criteria like ``our method'' and ``our usual means of investigation'' relevant to defining coarse-grained entropy? In the absence of an answer from them, we can turn to the review of \emph{Statistical Mechanics} by Samuel Burbury. 

Burbury's ``Definition A'' of density in phase \cite[p.~254]{burbury_variation_1903} corresponds to the limit in which the dimensions constituting $\delta $ ``become infinitely small.'' This Definition A thus corresponds to fine-grained entropy. Burbury later stated, ``Definition A {\dots} has the advantage of being mathematically irreproachable.'' That is a point with which Zermelo would firmly agree. Burbury, in contrast, added that it was ``for the present purpose, I think, useless.'' Burbury, then, in contrast to Gibbs and Poincaré proposed substituting, ``for the continuous liquid of Prof.~Gibbs's illustration, a medium consisting of discrete molecules, each of finite dimensions.'' The number of molecules within $\delta $ as it becomes infinitely small ``has no meaning whatever.'' In dealing with molecules, Definition A, and fine-grained entropy along with it, must be replaced.

The added consideration of discrete molecules led Burbury to propose ``a rough makeshift'' \cite[p.~257]{burbury_variation_1903}, the density at a point is the number of molecules within a sphere of finite radius centered on that point. The density in phase space then had an analogous definition. While the size of the sphere was still left unspecified, it then had an objective physical basis: molecular dimensions, or more generally the dimensions of the bodies in the specific application, e.g. asteroids, planets or stars. Those physical dimensions demand a finite size of $\delta $ and provide a physical justification for coarse-grained entropy.

Poincaré \cite[p.~373--4]{poincare_reflexions_1906-2} declared that coarse-grained entropy is what is usually conceived of in physics. He, further, proved two statements. Fine-grained entropy is constant over time; coarse-grained entropy is not. Coarse-grained entropy (supplying, for clarity, the factor $-k$ that Poincaré omits) is always larger than fine-grained entropy. Poincaré also stated here, but without proof, that the ``coarse entropy of physicists'' always increases.

There is nothing in Gibbs's discussion of coloring or Burbury's revised definition of density that precludes reversibility.

\section{ Discussion of Poincaré's Concepts}

In summary, we focused on two concepts about many-body dynamical systems found in Poincaré's 1906 article.

\textbf{The concept of fine and coarse graining }is due to Gibbs and he is generally correctly credited. The terms were however first used by Poincaré and it seems he never gets credit. Gibbs and Poincaré both attribute the need for coarse graining to limits on the fineness of our senses. Burbury, agreeing with the introduction of coarse graining by Gibbs, attributes the need for coarse graining instead to the physical dimensions of the bodies under consideration. He thus provides a physical basis (and not mathematical or philosophical) for the need for coarse graining. 

As has been justifiably emphasized, the fine-grained entropy of a Hamiltonian dynamical system is constant. This means that with fine graining, all Hamiltonian dynamical systems are fully reversible, and are not irreversible. The introduction of coarse graining offers an interpretation in which dynamical systems are not reversible but does not constrain the direction in which they can pass between two accessible states with equal energy. As already noted, it does not limit the entropy to only increasing or decreasing.

\textbf{The concept of uncertainty of initial conditions} was introduced to statistical mechanics by Poincaré in his 1906 paper. There he reasoned based on comparing two hypotheses. In one hypothesis, with an assumption of perfect knowledge of the initial conditions of a system, it follows that, because of the deterministic equations of motion, the entropy at any subsequent time is infinite. Poincaré rejects this conclusion and the assumption that led to it. He therefore accepts the hypothesis that our knowledge of the initial conditions and the subsequent conditions of the system are uncertain. 

In his 1890 work on dynamical systems, Poincaré showed that very small variations in the initial conditions in phase space of a system can result in substantially different trajectories; no two systems are truly identical. The immediate consequence of this conclusion is that ensembles of systems and thus ensemble averaging and statistics in general are needed and this is true of both planets and molecules. Further, because of this inherent uncertainty, there is no genuine \emph{a priori} outcome for a system. It also indicates an ontic frequentist understanding of probability in statistical mechanics is required.

\section{ Conclusion}

In summary, we focused on two concepts about many-body Hamiltonian dynamical systems found in Poincaré's 1906 article and important to the foundations of statistical mechanics. Poincaré's earlier work on dynamical systems makes possible the analytical mechanics is Gibbsian statistical mechanics.

\textbf{Coarse graining}, which Gibbs and Poincaré understood as demanded by the limits of our perception (like finding a grain of barley in a sack of wheat) and which Samuel Burbury connected with atomic and molecular dimensions, allows the possibility of a state variable which changes. It does not set the direction for the change, increasing or decreasing.

\textbf{Uncertain initial conditions} also affect the very-many-body systems of statistical mechanics. Poincaré in 1890 showed that for three-body systems governed by dynamical, analytic equations of mechanics, nearly identical initial conditions (near a homoclinic point) may result in very different trajectories in phase space. 

Although, in Gibbsian statistical mechanics as amplified by Poincaré, the equations of motion are deterministic, ensemble averages of systems are needed because the trajectories of systems, and their stability or instability, are acutely sensitive to initial conditions. Multiple instantiations of one dynamical system (with nearly identical initial conditions) could follow vastly different trajectories. It follows from this that the use of phase averages is justified, even required. The relation between this understanding of phase average and time average was not considered. Further, phase averages give results matching experiments. The thin separation between initial conditions leading to stable or unstable trajectories means that ensemble probability is ontic and frequentist and does not have an \emph{a priori} value. 

The idea of identifying non-deterministic behavior of dynamical systems as a source of irreversibility seems both attractive and hard to argue persuasively.

\backmatter

\bmhead{Conflict of Interest Statement} I have no relevant financial or non-financial interests to disclose.

I did not receive support from any organization for the submitted work.

\bibliography{Poincare_and_Gibbs}

\end{document}